\documentclass[12pt]{article}    

\usepackage[left=1.3in,top=1.5in,right=1.3in,bottom=1.25in]{geometry}
\usepackage{setspace}
\usepackage[normalem]{ulem}
\usepackage{amsfonts}
\usepackage{amssymb}
\usepackage{epsfig}
\usepackage{amsmath}
\usepackage[numbers]{natbib}

\newcommand{\bs}{\boldsymbol}
\newcommand{\bd}{\bf}
\newcommand{\dist}{\mathsf}
\newcommand{\ep}{\epsilon}
\newcommand{\1}{\mathbf{1}}

\newcommand{\beff}{\mathbf{f}}

\title{The partition problem: case studies in Bayesian screening for time-varying model structure}

\author{Zesong Liu \\
Jesse Windle \\
James G.~Scott \\
\\
The University of Texas at Austin}

\begin{document}

\begin{spacing}{1.05}

\maketitle

\begin{abstract}

This paper presents two case studies of data sets where the main inferential goal is to characterize time-varying patterns in model structure.  Both of these examples are seen to be general cases of the so-called ``partition problem,'' where auxiliary information (in this case, time) defines a partition over sample space, and where different models hold for each element of the partition.  In the first case study, we identify time-varying graphical structure in the covariance matrix of asset returns from major European equity indices from 2006--2010.  This structure has important implications for quantifying the notion of financial contagion, a term often mentioned in the context of the European sovereign debt crisis of this period.  In the second case study, we screen a large database of historical corporate performance in order to identify specific firms with impressively good (or bad) streaks of performance.

\vspace{0.1in}

\end{abstract}

\newpage

\section{Introduction}

Problems of model selection are often thought to be among the most intractable in modern Bayesian inference.  They may involve difficult high-dimensional integrals, nonconcave solution surfaces, or large discrete spaces that cannot be enumerated.  All of these traits pose notorious computational and conceptual hurdles.  

Yet model-choice problems---particularly those related to feature selection, large-scale simultaneous testing, and inference of topological network structure---are also some of the most important.  As modern data sets have become larger, they have also become \textit{denser}---that is, richer with covariates, more deeply layered with underlying patterns, and indexed in ever more baroque ways (e.g.~$x_{ijk t}$, rather than just $x_{ij}$).  It is this complex structure, more than the mere tallying of terabytes, that defines the new normal in 21st-century statistical science.  There is thus a critical need for Bayesian methodology that addresses the challenges posed by such data sets, which come with an especially compelling built-in case for sparsity.

The difficulties of model selection are further exacerbated when the structure of the model is not ``merely'' unknown, but also changes as a function of auxiliary information.  We call this the \textit{partition problem}: the auxiliary information defines an unknown partition over sample space, with different unknown models obtaining within different elements of the partition.  Here are three examples where model structure plausibly covaries with external predictors.
\begin{description}
\item[Subgroup analysis:] How can clinicians confront the multiplicity problem inherent in deciding whether a new cancer drug is effective for a specific subgroup of patients, even if it fails for the larger population?  Here the model is simply a binary indicator for treatment effectiveness, while the partitions are defined by diagnostically relevant covariates---for example, age, sex, or smoking status.  No good approaches exist, Bayesian or otherwise, that are capable of systematically addressing this problem.  The difficulty is that examining all possible partitions wastes power: many substantively meaningless or nonsensical partitions are considered, and must receive prior probability at the expense of the partitions we care about.
\item[Partitioned variable selection:] A patient comes to hospital complaining of a migraine.  The hospital wishes to use the patient's clinical history to diagnose whether the migraine may portend a subarchnoid hemorrhage, a catastrophic form of brain bleed.  Such hemorrhages are thought to be etiologically distinct for children and adults.  Thus the age of the patient influences which aspects of her clinical history (i.e.~variables) should be included in the predictive model.
\item[Network drift:] A delay-tolerant network (DTN) for communication devices has few instantaneous origin-to-destination paths.  Instead, most messages are passed to their destination via a series of local steps (at close range) from device to device.  In such a setting, it helps to know the underlying social network of users in the model---that is, who interacts with whom, and how often---in order to predict the likelihood of success for specific directed transmissions.  Thus the time-varying topological structure of the \textit{social} network has important implications for the efficient routing of traffic within the \textit{device} network.
\end{description}

All three of these problems recall the literature on tree modeling, including \citep{breiman:friedman:etal:1984}, \citep{chipman:george:mcculloch:1998}, \citep{denison:mallick:smith:1998}, and \citep{gramacy:lee:2008}.  Yet to our knowledge no one has studied tree models wherein one large discrete space (for example, predictors in or out of a linear model) is wrapped inside another large discrete space (trees). 

This poses all the usual computational and modeling difficulties associated with tree structures, and large discrete spaces more generally.  But it also poses a major, unique challenge.  In all existing applications of Bayesian tree models we have encountered, the collapsed sampler (whereby node-level parameters are marginalized away and MCMC moves are made exclusively in tree space) is the computational tool of choice.  But in the partition problem, the bottom level parameter in the terminal nodes of the tree is itself a model indicator, denoting an element of some potentially large discrete space.  This makes it difficult to compute the marginal likelihood of a particular tree in closed form, since doing so would involve a sum with too many terms.  The collapsed sampler therefore cannot always be used.

Trees are, of course, just one possible generative model for partitions of a sample space based on such auxiliary information.  Others include species-sampling models, coalescent models, or urn models (such as those that lie at the heart of Bayesian nonparametrics).  But no matter what partitioning model is entertained, new models and algorithms are necessary to make inferences and quantify uncertainty for this very general class of problems.

This chapter presents two case studies of data sets within this class.  For both data sets the auxiliary information is time.  In the first case study, we identify time-varying graphical structure in the covariance of asset returns from major European equity indices from 2006--2010.  This structure has important implications for quantifying the notion of financial contagion, a term often mentioned in the context of the European sovereign debt crisis of this period.  In the second case study, we screen a large database of historical corporate performance in order to identify specific firms with impressively good (or bad) streaks of performance.

Our goal in these analyses is not to address all the substantive issues raised by each data set.  Rather, we intend: (1) to present our argument for the existence of non-trivial dynamic structure in each data set, an argument that can be made using simple models;  (2) to draw parallels between the case studies, both of which exemplify the partition problem quite well; and (3) to identify certain aspects of each model that must be generalized in future work if these case studies are to provide a useful template for other data sets.

\section{Case study I: financial contagion and dynamic graphical structure}

\subsection{Overview}

During times of financial crisis, such as the bursting of the US housing bubble in 2008 and the European sovereign-debt crisis in 2010, the co-movement of asset prices across global markets is hypothesized to diverge from its usual pattern. Many financial theorists predict, and many empiricists have documented, changes in market relationships after these large market shocks. It is important to track these large shocks and measure their impacts on the global economy so that we can better understand future market behaviors during times of crisis.

In general, the idea that market relationships change after large shocks is called contagion \citep{dornbusch:2000}.  In the literature, there has been a lengthy debate over precise definition of this term.  As a practical matter, we define contagion as significant change in the pattern of correlation in the residuals from an asset-pricing model during times of crisis, following in the tradition of previous authors \citep[e.g.][]{forbes:rigobon:2002,bae:karolyi:stutz:2003,bekaert:harvey:ng:2005}. Focusing primarily on how the relationships between markets change, we want to determine whether large shocks have significant impact on the subsequent interactions between markets.

The standard way to study the co-movements and interdependent behavior of markets is by looking at the covariance matrices of returns across different countries. In this case study, we explore ways of estimating this covariance structure to study the change of the market dynamic over time. Normally, when constructing covariance matrices, the algorithms applied are computationally identical to repeated applications of least squares regressions.  Instead, we apply the ideas of Bayesian model selection, using the Bayesian information criterion, or BIC \citep{schwarz1978}, to approximate the marginal likelihoods of different hypotheses, and a flat prior over model space.

In applying this method, we uncover many signs of contagion, which manifests itself as time-varying graphical structural in the covariance matrix of returns. For example, if we look at the relationship between Italy and Germany, the traditionally positive correlation between the countries changes sign  during the sovereign debt crisis. This provides just one example of the evidence for contagion discovered in these investigations.

\subsection{Contagion in factor asset-pricing models}

The sovereign debt crisis started when Greece became in danger of defaulting on its debt. For years, Greece had been a rapidly growing economy with many foreign investors. This strong economy was able to withstand large government deficits that Greece had during that time. But after the worldwide 2008 financial crisis hit, two of the country's largest industries, tourism and shopping, were badly affected. This downturn caused panic in the Greek economy. Although Greece was not the only country that confronted debt problems, its debt-to-GDP ratio was judged excessively high by markets and ratings agencies, reaching 120\% in 2010.  Moreover, one of the major fears that arose during this period was that investors would lose faith in other similary situated Euro-zone economies, which could cause something similar to a run on a bank.

One of the major events in this episode occurred on May 9, 2010. On that day the 27 member states of the EU agreed to create the European Financial Stability Facility, a legal instrument aimed at preserving financial stability in Europe by providing financial assistance to states in need. This legislation was upsetting to countries with large healthy economies such as Germany, whose electorate focused on the negative effects of the bailout.  In light of these developments, not only do we want to show that the pattern of asset-price correlation changed, but we also want to make sense of these changes by linking them to the news headlines about bailouts.

Our raw data are daily market returns from equity indices corresponding to nine large European economies---Germany, the UK, Italy, Spain, France, Switzerland, Sweden, Belgium, and the Netherlands---from December 2005 to October 2010.  We do not include Greece because of its small size relative to the other economies of Europe,  but we do include the Euro--Dollar exchange rate as a tenth column in the data set.

By our definition of contagion, we need to examine the residuals of the returns within the context of an asset-pricing model. Specifically, we use a four-factor model where
$$
E(y_{it} \mid EU , US ) = \beta_i^{US} x^{US}_{t} + \beta_i^{EU} x^{EU}_{t} + \gamma_i^{US} \delta^{US}_{t} + \gamma_i^{EU} \delta^{EU}_{t} \, ,
$$
where $y_{it}$ is the daily excess return on index $i$; $x^{US}_{t}$ is the daily excess return on a value-weighted portfolio of all US equities; $x^{EU}_{t}$ is the daily excess return on the EU-wide index of Morgan Stanley Capital International; $\delta^{US}_{t}$ is the volatility shock to the US market; and $\delta^{EU}_{t}$ is the excess volatility shock to the European market.  The excess shock is defined as the residual after regressing the EU volatility shock upon the US volatility shock.  This is necessary to avoid marked collinearity, since the US volatility shock strongly predicts the EU volatility shock.  These volatility factors are calculated using the particle-learning method from \citep{Polson:Scott:2011e}, not described here.

In this model, the loadings $\beta_i^{US}$ and $ \beta_i^{EU}$ measure the usual betas relative to the US and Europe-wide equity markets.  Thus we have controlled for regional and global market integration via an international CAPM-style model \citep{bekaert:harvey:ng:2005}.  The loadings $\gamma_i^{US}$ and $\gamma_i^{EU}$, meanwhile, measure country-level dependence upon global and regional volatility risk factors.  As shown in \citet{Polson:Scott:2011e}, these loadings can be interpreted in the context of a joint model that postulates correlation between shocks to aggregate market volatility and shocks to contemporaneous country-level returns.

\subsection{A graphical model for the residuals}

The above model can be fit using ordinary least squares, leading to an estimate of all model parameters along with a set of residuals $\epsilon_{it}$ for all indices.  We now turn to the problem of imposing \textit{graphical} restrictions on the covariance matrix of these residuals.

A Gaussian graphical model defines a set of pairwise conditional-independence relationships on a $p$-dimensional zero-mean, normally distributed random vector (here denoted $x$).  The unknown covariance matrix $\Sigma$ is restricted by its Markov properties; given $\Omega = \Sigma^{-1}$, elements $x_i$ and $x_j$ of the vector $x$ are conditionally independent, given their neighbors, if and only if $\Omega_{ij} = 0$.  If $G=(V,E)$ is an undirected graph whose nodes represent the individual components of the vector $x$, then $\Omega_{ij} = 0$ for all pairs $(i,j) \notin E$.  The covariance matrix $\Sigma$ is in $M^+(G)$, the set of all symmetric positive-definite matrices having elements in $\Sigma^{-1}$ set to zero for all $(i,j) \notin E$.

We construct a graph by selecting a sparse regression model for each country's residuals in terms of all the other countries, as in \citep{Dobra04JMVA} and \citep{scottcarvalho2007b}.  We then cobble together the resulting set of conditional relationships into a graph to yield a valid joint distribution.  Each sparse regression model is selected by enumerating all $2^9$ possible models, and choosing the one that minimizes the BIC.  This leads to a potentially sparse model for $E(\epsilon_{it} \mid \epsilon_{j,t}, j \neq i)$.

In this manner, an adjacency matrix can be constructed for the residuals from the four-factor model.  The $(i,j)$ element of the adjacency matrix is equal to 1 if the residuals for countries $i$ and $j$ both appear in each other's conditional regression models, and is equal to 0 otherwise.  One may also assemble the pattern of coefficients from these sparse regressions to reconstruct the covariance matrix for all the residuals, denoted $\Sigma$.

We actually look at adjacency matrices and covariance matrices on a rolling basis, since we believe that there are changes in $\Sigma$ over time.  Each window involves a separate set of regressions for a period of 150 trading days, which is thirty weeks of trading, or about 7 months. We shift the window in 5 day increments, thereby spanning the whole five-year time period in our study.  For each 150-day window, we refit the estimate for $\Sigma_t$ using the entire graphical-model selection procedure.  For the sake of comparison, we also include the estimates of $\Sigma_t$ using OLS, ridge regression, and lasso regression.

\subsection{Results}

\begin{figure}
\begin{center}
\includegraphics[width=4.5in]{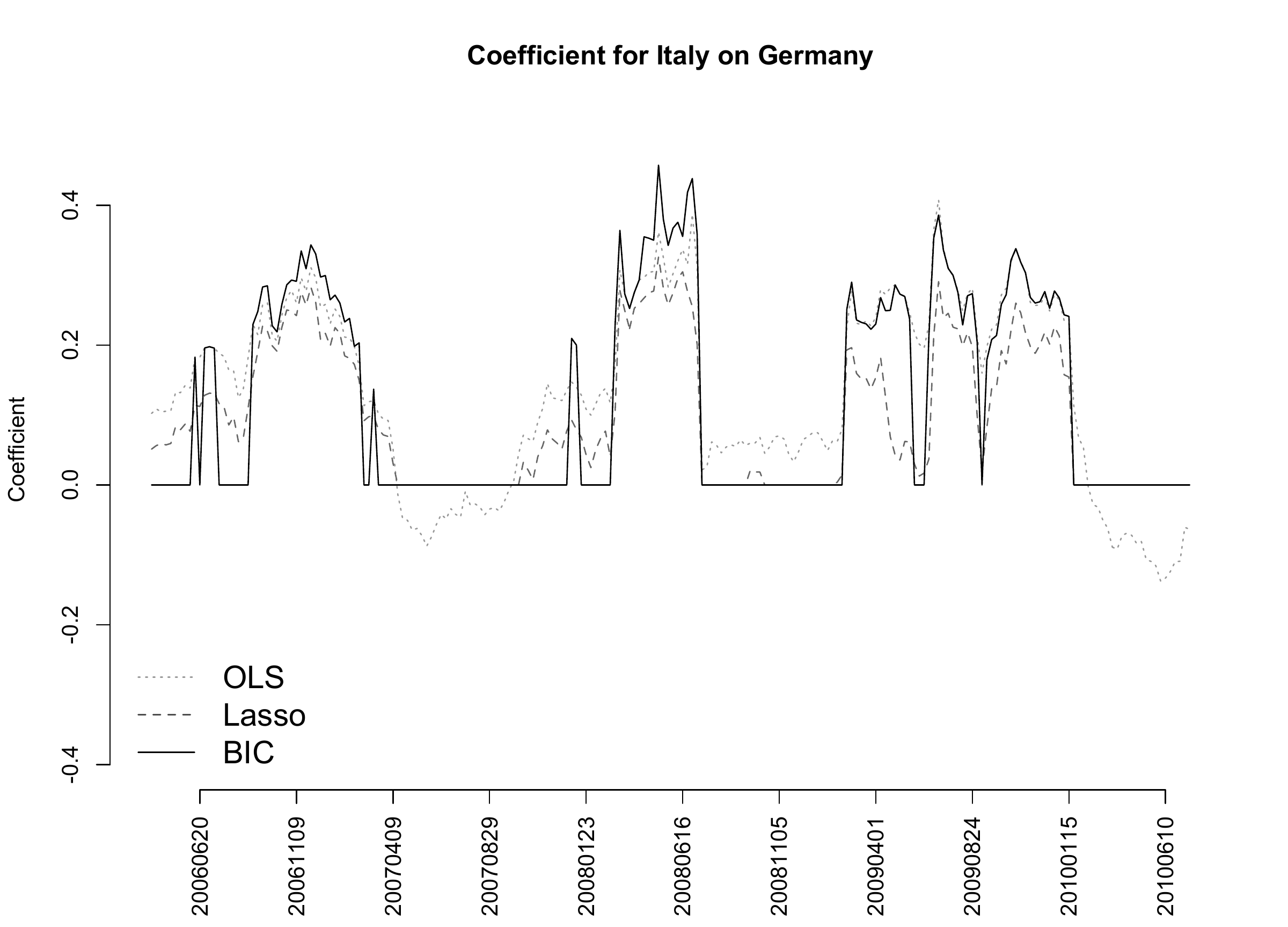} \\
\vspace{2pc}
\includegraphics[width=4.5in]{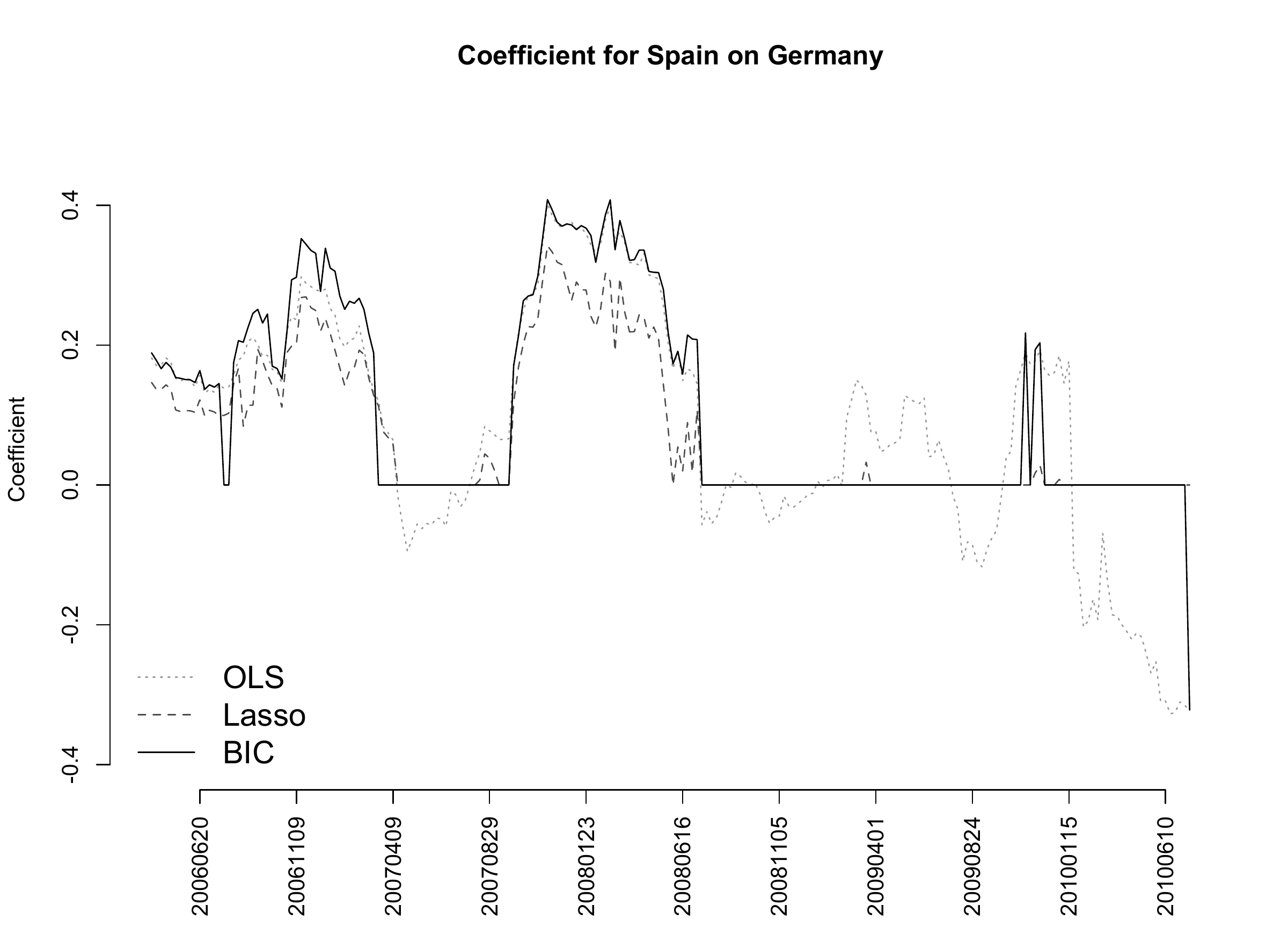} 
\caption{\label{fig:DEUresults} Estimated regression coefficients for Italy (top) and Spain (bottom) on Germany, where the coefficients are calculated on a rolling basis using Bayesian model selection via BIC.}
\end{center}
\end{figure}

\begin{figure}
\begin{center}
\includegraphics[width=4.5in]{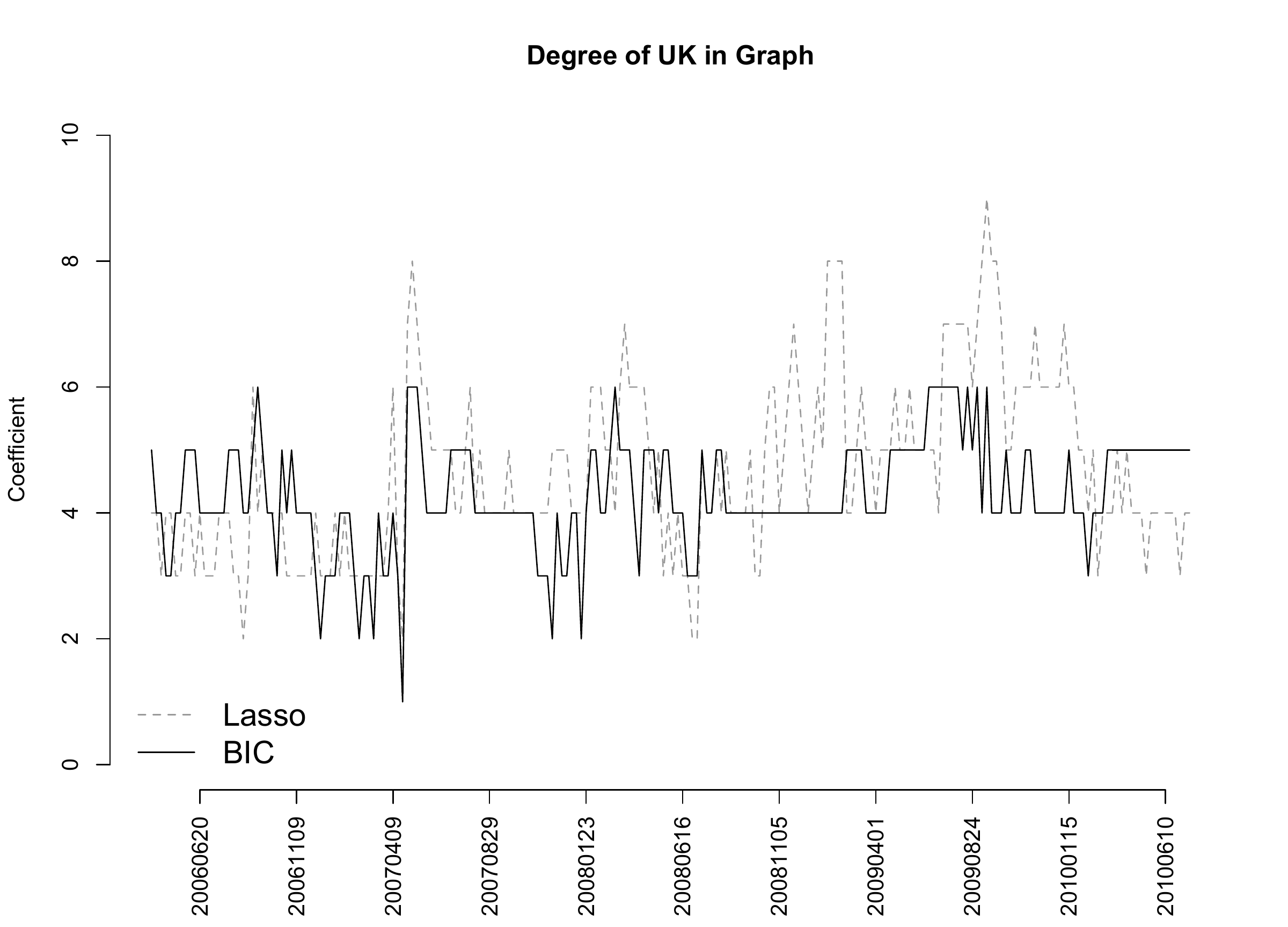} \\
\vspace{2pc}
\includegraphics[width=4.5in]{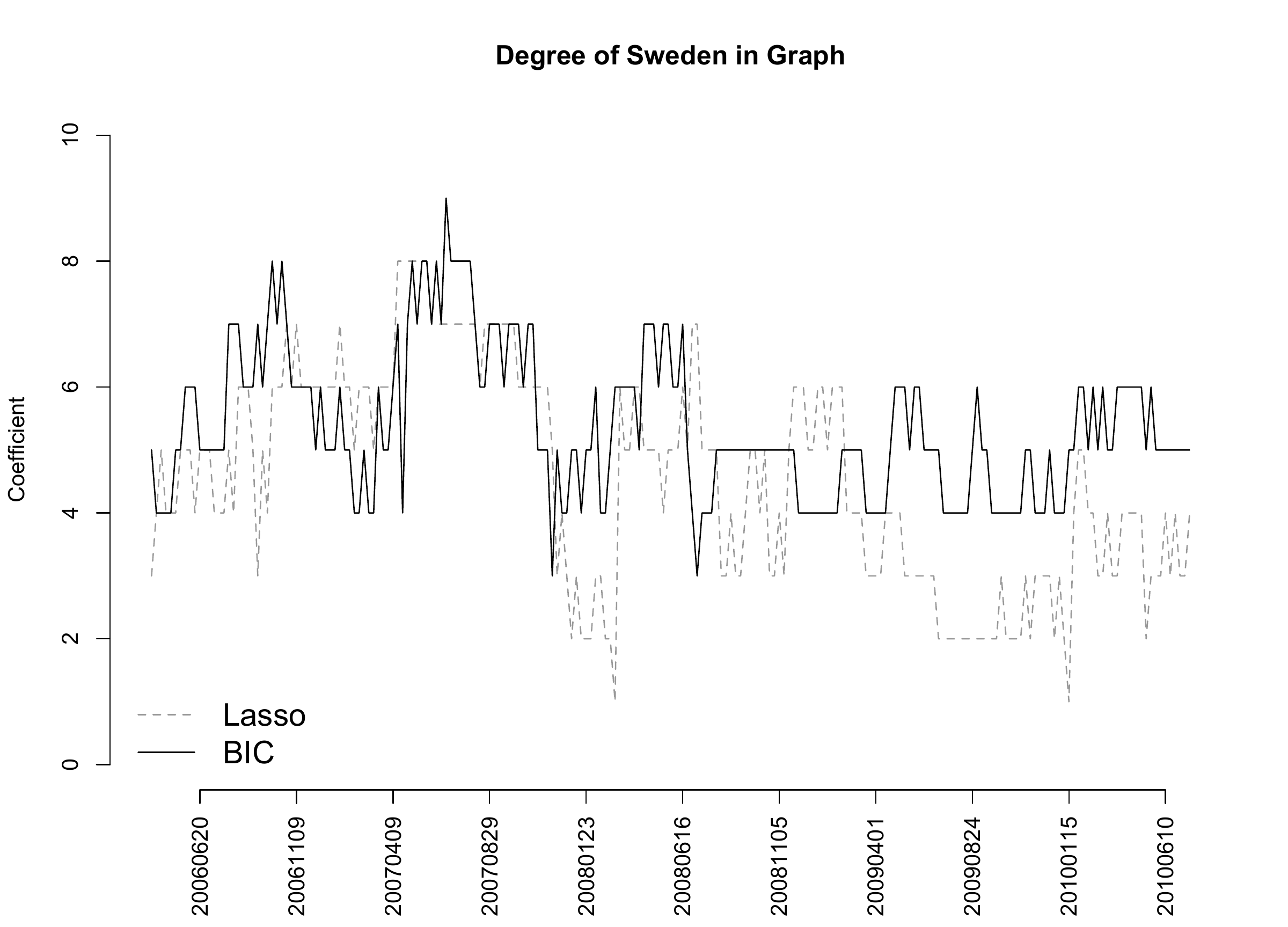} 
\caption{\label{fig:adjacency} Estimated adjacency degree in the time-varying graph of residuals for the UK (top) and Sweden (bottom).  There seems to be clear evidence of time-varying topological structure in the graph.}
\end{center}
\end{figure}

From Figure \ref{fig:DEUresults}, which depicts the rolling estimates of the Italy--Germany and Spain--Germany regression coefficients, it is clear that there are nonrandom patterns that remain in the residuals. If the factor model fully explained the co-movements in the European market, we would expect the residuals to have no covariance and look like noise. We would also expect that the Bayesian variable-selection procedure would give regression coefficients of 0. To be sure, the data support the hypothesis that specific elements of the precision matrix $\Omega_t = \Sigma_t^{-1}$ are zero over certain time periods.  An example of this is the ITA-DEU coefficient during much of 2008.  Yet it is patently not the case that all such elements are zero for all time periods: the standard significance test of an empty graph is summarily rejected ($p<10^{-6}$) at all time points.  For details of this test, see \citep{lauritzen96} and Proposition 1 of \citep{carvalhoscott2007}.

There are several explanations for the observed correlation between the residuals. First, it is highly probable that the factor models are imperfect. Regressing only on the market returns and market volatility, the four factor model involves a substantial simplification of reality.  For example, during the sovereign debt crisis, we could reasonably add an explanatory variable that takes into account the change in likelihood of a bailout. Second, even if we imagine that the four-factor model is the true model for this system of markets, we only have proxies for both the global and local returns and volatility. Specifically, using the US market return as a proxy for the world market return is a reasonable estimate, but is far from perfect.  Moreover, the factors that measure market volatility are at best a measure of the average market volatility over a short time span.  This could potentially distort the residuals, since we cannot observe volatility spikes on a more granular scale.  Most likely, the correlations in the residuals stems from some mixture of these two effects.

The same conclusion about significant residual correlation is also borne out by examining the time-varying topology of the graph itself.  While it is difficult to visualize the time-varying nature of the network's structure in its entirety, we can look at quantities such as how the adjacency of a specific node in the graph---that is, how many neighbors it has---changes over time.  Figure \ref{fig:adjacency} shows the estimated adjaceny degrees for Sweden and the UK, two non-Euro countries.  Again we see that the factor model is not perfect, as the residuals still exhibit correlation.  On the other hand, we see that the degree of each vertex is not 9 at every time point. This means that shrinkage is often useful: it is clear that estimating $p(p-1)/2$ separate correlation parameters is a poor use of the data, and will lead to estimators with unnecessarily high variance.  We can almost certainly reduce the required number of parameters while still obtaining a good estimate.  This illustrates the utility of the graphical modeling approach.

Finally, the relationship between the Spain and Germany residuals is easily interpretable in terms of the underlying economic picture.  In the summer of 2010, there is an apparent divergence from the historical norm, precisely coinciding with the Greek sovereign-debt crisis and associated bailout.  The historically aberrant negative correlations between the residuals from Germany and the southern countries suggests that markets reacted very differently in these two countries to news of the period.  A useful comparison is with the period of September and October 2008, when the global financial crisis associated with the bursting of the housing bubble was at its peak.  These were global rather than EU-centric events, and no such divergence was observed involving the German-market residuals.

Our approach provides initial evidence for contagion effects, but has some important limitations.  In particular, we have estimated time-varying graphical structure using a moving-window variable selection approach, which does not explicitly involve a dynamic model over graph space.  Some authors have made initial efforts in studying dynamic graphs \citep[e.g.][]{taddy:gramacy:polson:2011,wang:reeson:carvalho:2011}, but much further work remains to be done to operationalize the notion of contagion within this framework.  The issue is that we expect contagion to be associated with a sharp change in the underlying graphical structure of the residual covariance matrix, as opposed to the locally drifting models considered by these other authors.  Such sharp changes are likely obscured by our rolling-regression approach, in that only about $3\%$ of the data changes in each new window.

\section{Case study II: simultaneous change-point screening and corporate out-performance}

\subsection{Overview}

In this case study, we compare publicly traded firms against their
peer groups using a standard accounting metric known as ROA, or return
on assets. The data set comprises 645,456 company-year records from
53,038 companies in 93 different countries, spanning the years
1966-2008.

Just as in the previous example, we will attempt to uncover substantively meaningful changes over time in the underlying model for each time series.   Let $y_{it}$ denote the benchmarked ROA observation for company $i$ at
time $t$; let $\bd{y}_i$ denote the whole vector of observations for
company $i$ observed at times $\bd{t}_i = (t_1,...,t_{n_i})$; and let
$\bd{Y}$ denote the set of $\bd{y}_i$ for all $i$.  We say the observations have been ``benchmarked'' to indicate that they have undergone a pre-processing step that removes the effects of a firm's size, industry, and capital structure.  For details, see \citep{scott:2010b}.

The goal is to categorize each time series $i$ as either signal or
non-signal, both to be defined shortly.  The model we consider takes the general form
\begin{align}
\bd{y}_i &= \bd{f}_i + \bs{\ep}_i, \; \bs{\ep}_i \sim \dist{Noise} \\
\bd{f}_i &\sim \omega \cdot \dist{Signal} + (1-\omega) \cdot \dist{Null}
\end{align}
where $\dist{Signal}$ is the distribution describing the signal and $\dist{Null}$ is the
distribution describing the non-signal.  In this case,
``noise'' should not be conflated with actual measurement
error. Instead, it represents short-term fluctuations in performance
that are not indicative of any longer-term trends, and are thus not of particular relevance for understanding systematic out-performance.

We can rephrase this model as
\begin{align}
\bd{y}_i &= \bd{f}_i + \bs{\ep}_i, \; \bs{\ep}_i \sim \dist{Noise} \nonumber \\
\bd{f}_i &\sim \dist{F}_{\gamma_i}, \;
         \dist{F}_1 = \dist{Signal}, \; \dist{F}_0 = \dist{Null} \label{Model1}  \\
\gamma_i &\sim \dist{Bernoulli}(\omega) \nonumber
\end{align}
which provides us with the auxiliary variable $\gamma_i$ that
determines whether a time series $i$ is either signal or noise.  Thus the posterior distribution $p(\gamma_i=1 | \bd{Y})$ is a
measure of how likely time series $i$ is signal.  One can sort the
data points from most probably to least probably signal by simply
ranking $p(\gamma_i=1 | \bd{Y})$.  Importantly, these posterior probabilities will contain an automatic penalty for data dredging: as more unimpressive firms are thrown into the cohort, the posterior distribution for $\omega$ will favor increasingly smaller values, meaning that all observations have to work harder to overcome the prior bias in favor of the null \citep[see, e.g.,][]{scottberger06,bogdan:ghosh:2008b}.

Suppose, for example, that $\bd{f}_i = 0$ corresponds to the average performance of a company's peer group, and that we want deviations from zero to capture long-range fluctuations in $\bd{y}_i$ from this peer-group average---that is, changes in a company's fortunes or long-run trends that
unfold over many years.  Several previous authors have proposed methodology for the multiple-testing problem that arises in deciding whether $\beff_i = 0$ for all firms simultaneously \citep{Scott:2008,henderson:etal:2009,polson:scott:2009a}.  The origin of such a testing problem lies in the so-called ``corporate success study,'' very popular in the business world: begin with a population of firms, identify the successful ones, and then look for reproducible behaviors or business practices that explain their success.

The hypothesis that $\bd{f}_i = 0$, then, implies that a firm is no better,
or worse, than its peer group over time. One way to test this
hypothesis is by placing a Gaussian-process prior on those $\bd{f}_i$'s
that differ from zero. Suppose, for example, we have an i.i.d. error
structure and some common prior inclusion probability:
\begin{align*}
\bd{y}_i &= \bd{f}_i + \bs{\ep}_i, \; \bs{\ep_i} \sim \dist{N}(0,\sigma^2 I) \\
\bd{f}_i &\sim \dist{F}_{\gamma_i}, \;
  \dist{F}_1 = \dist{N}(0, \sigma^2 K(\bd{t_i})), \; \dist{F}_0 = \delta_0 \\
\gamma_i &\sim \dist{Bernoulli}(\omega)
\end{align*}
where $K(\bd{t})$ is the matrix produced by a covariance kernel
$k(\cdot,\cdot)$, evaluated at the observed times $\bd{t} =
(t_1,...,t_M)'$. The $(i,j)$ entry of $K$ is:
\[
K_{i,j} =k(t_i,t_j).
\]
Typically the covariance function will itself have hyperparameters
that must be either fixed or estimated.  A natural choice here is the squared-exponential kernel:
\[
k(t_i,t_j)= \kappa_1 \exp \Big\{-\frac{(t_i-t_j)^2}{2 \kappa} \Big\} +
\kappa_3 \delta_{t_i,t_j}
\]
where $\delta_{t_i,t_j}$ is the Kronecker delta function. If the three
$\kappa$ hyperparameters are fixed, then this is just a special case
of a conjugate multivariate Gaussian model, and the computation of the
posterior model probabilities $p(\gamma_i=1 | \bd{Y})$ may proceed by
Gibbs sampling.  As shown in \citep{Scott:2008}, this basic idea may be generalized
to more complicated models involving autoregressive error structure and Dirichlet-process priors on nonzero trajectories.

\subsection{Detecting regime changes}

One shortcoming of this strategy is that $\beff_i$ is assumed to be either globally zero or globally nonzero.  The partition problem is different, and captures an important aspect of reality ignored by the model above: the signal of interest may not involve
consistent performance, but rather a precipitous rise or fall in the
fortunes of a company.

We therefore consider the possibility that each firm's history may, though not necessarily, be
divided into two epochs.  The separation of these epochs corresponds
to some sort of significant schism in performance between time periods.
For instance, a positive jump might arise by virtue or a drug patent
or the tenure of an especially effective leader--someone like Steve
Jobs of Apple, or Jack Welch of General Electric. There may also be
periods of inferior performance when the jump is negative.

We therefore adapt model (\ref{Model1}) in the following way.  With each firm, we associate not a binary indicator of ``signal'' or ``noise'', but rather a multinomial indicator for the location in time of a major shift in performance.  We then index all further parameters in the model by this multinomial indicator:
\begin{align}
\bd{y}_i &= \bd{f}_i + \bs{\ep}_i, \; \ep_i \sim \dist{Noise} \nonumber \\
\bd{f}_i &\sim \dist{F}_{\gamma_i} \label{Model2} \\
\gamma_i &\sim \dist{Multinomial}(\bs{\omega}) \nonumber \, ,
\end{align}
with the convention that $\gamma_i = 0$ denotes the no-split case where $\beff_i$ is globally zero, and $\gamma_i = n$ the no-split case where $\beff_i$ is globally non-zero.

This differs from the traditional changepoint-detection problem in two respects: we are data-poor in the time direction, but data-rich in the cross-section.  Indeed, this case study is the mirror image of the previous one, where the number of time series was moderate the number of observations per times series was large.  This fact requires us to consider models that are simpler in the time domain, but also allows us to borrow cross-sectional information across firms for the purpose of estimating shared model parameters.  This is particularly important for the multinomial parameter $\bs{\omega}$, which lives on the $(n+1)$-dimensional simplex, and descibes the population-level distribution of changepoints.

There are many potential choices for $\dist{F}_k$ and $\dist{Noise}$.
We instance, we could choose
\[
F_k = \dist{N}(0, C_k) \textmd{ and } \dist{Noise} = \dist{N}(0, \Sigma)
\]
where $\Sigma$ describes the covariance structure of the noise and $C_k$ describes a split in epochs at time
$k$ (again recalling the convention that $C_0$ is degenerate at zero and that $C_n$
corresponds to no split at all).  Of course, we need not limit ourselves
to this interpretation, and may intead choose a collection of $\{C_k\}$
that embodies some other substantive meaning.  For instance, we could consider the
collection $C_{i,k}$ where $C_{1,k}$ corresponds to a small jump at
time $k$ and $C_{2,k}$ corresponds to a large jump at time $k$.  We could also generalize to multiple regime changes per firm.  For now, however, we consider the simpler case where there can be at most one shift, and where all shifts are exchangeable.  Recall that we intend such an intentionally oversimplified model to be useful for high-dimensional screening, not nuanced modeling of an individual firm's history.

As before, both
$\{C_k\}$ and $\Sigma$ will include some sort of
hyperparameters, which we will denote as $\bs{\theta}$ for now.
Conditional on the hyperparameters, we may write the distribution of the data as
\[
\bd{y}_i | \gamma_i \sim \dist{N}(0, C_{\gamma_i} + \Sigma),
\]
which are conditionally independent across $i$.  Thus when calculating
the posterior distribution using a Gibbs sampler, the conditional distribution
$p(\bs{\gamma} | Y, \bs{\theta}, \bs{\omega})$ will conveniently
decompose into a product over $p(\gamma_i | \bd{y}_i, \bs{\theta},
\bs{\omega})$.  Moreover, when each time series has its own hyperparameter $\bs{\theta}_i$ and
the only shared information across time series is the multinomial probability vector
$\bs{\omega}$, the posterior calculation simplifies further.  In
particular, after marginalizing out each $\bs{\theta}_i$, the only quantities that must be sampled are $\{ p(\gamma_i | \bd{y}_i, \bs{\omega}) \}$ and
$p(\bs{\omega} | \bd{Y}, \bs{\gamma})$.

Much of the details of such a model are encoded in particular choices for the covariance matrices describing signals and noise.  As an illustration of this general approach, consider a simple model in which $\bd{f}_i$ is piecewise
constant.  This assumption is reasonable given the relatively short
length of each time series; in any case, one may think of it as a
locally constant approximation to the true model.  Suppressing the index $i$ for the moment, we write the model as
\begin{align*}
y_{\bd{s}} &= f_{\bd{s}} + \ep_{\bd{s}}, \ep_{\bd{s}} \sim \dist{N}(0,
\sigma^2_{\bd{s}} I ), \\
f_{\bd{s}} &\equiv \theta, \\
\theta &\sim \dist{N}(0, \sigma^2_{\bd{s}} \tau^2), \\
\sigma^2_{\bd{s}} &\sim \dist{IG}(a/2, b/2)
\end{align*}
where $\bd{s}$ is some subsequence of the times $\{1, \ldots, n\}$.
Marginalizing over $\theta$ and $\sigma^2_{\bd{s}}$ yields a multivariate-T marginal:
\begin{align*}
y_{\bd{s}} &\sim \dist{T}_{a + |\bd{s}|}(0, R_{\bd{s}}) \\
R_{\bd{s}} &= \frac{a}{b} (I_{\bd{s}} + \tau^2 S_{\bd{s}}) \, .
\end{align*} 
If we know $\theta = 0$, or in other words that $\theta \sim \delta_0$, then \(
R_{\bd{s}} = (a/b) I_{\bd{s}}.  \) Notice that this formulation
automatically handles missing data, since one can simply exclude those
times from $\bd{s}$.

This can be phrased in terms of model (\ref{Model2}) by letting $S_k =
\1_k \1_k'$ be the $k \times k$ matrix of ones, and defining
\[
\dist{F}_k = \dist{N} (0, \tau C_k) \, ,
\]
where
\[
C_k | \sigma_{ij}^2 =
\begin{bmatrix}
\sigma^2_{i1} S_k & 0 \\
0 & \sigma^2_{i2} S_{n-k}
\end{bmatrix}
\textmd{ for } k = 1, \cdots, n-1,
\]
$\dist{F}_n = N(0, \sigma_i^2 \tau S_n)$, and $\dist{F}_0 = \delta_0$
(that is $C_n = \sigma^2_i S_n$ and $C_0 = 0$).  Furthermore, we
define the prior over the residuals so that it, too, depends on the index $\gamma_i$:
\[
\dist{Noise} = \dist{E}_{\gamma_i} = N(0, \Sigma_{\gamma_i})
\]
where $\Sigma_{k} | \sigma_{ij}^2 = \sigma_{1i}^2 I$ for $k = 0$ and
$n$, and where
\[
\Sigma_{k} | \sigma_{ij}^2 =
\begin{bmatrix}
\sigma_{i1}^2 I & 0 \\
0 & \sigma_{i2}^2 I
\end{bmatrix}
\textmd{ for } k = 1, \ldots, n-1 \, ,
\]
Finally, for $p(\bs{\omega})$ we assume a conjugate Dirichlet prior (details below).

This simple model has many advantages: it is
analytically tractable; it handles missing data easily; it allows for the possibility of a
sharp break between epochs; and it allows for preprocessing of all marginal likelihoods, which
saves time in the Gibbs sampler.  To see this, observe that the conditional
posterior distribution for $\bs{\gamma}$ is
\[
p(\bs{\gamma} | \bd{Y}, \bs{\omega}) \propto
\prod_{i} p(\bd{y}_i | \gamma_i, \bs{\omega}) p(\gamma_i | \bs{\omega})
\]
so we can sample each $p(\gamma_i | \bd{y}_i, \bs{\omega})$
independently.  In particular,
\[
p(\gamma_i = k | \bd{Y}, \bs{\omega}) \propto p(\bd{y}_i | \gamma_i =
k) p(\gamma_i = k | \bs{\omega}).
\]
Let $\ell_k$ be the observation times which are less than or equal
to $k$ and $r_k$ be the observation times which are greater than $k$,
both of which depend upon $i$.  When $k > 0$,
\[
p(\gamma_i = k | \bd{Y}, \bs{\omega}) \propto
\dist{T}_{a + |\ell_k|}(\bd{y}_i(\ell_k); 0, R_{\ell_k}) \, \cdot \,
\dist{T}_{a + |r_k|}(\bd{y}_i(r_k); 0,R_{r_k}) \, \cdot \,
p(\gamma_i = k | \bs{\omega})
\]
with the convention that $r_k = \emptyset$ if $k = n$.  When $k =
0$,
\[
p(\gamma_i = k | \bd{Y}, \bs{\omega}) \propto
\dist{T}_{a + |\ell_k|}(\bd{y}_i(\bd{t}_i); 0, \frac{a}{b} I_{\bd{t_i}}) \;
p(\gamma_i = k | \bs{\omega})
\]
All of the $\dist{T}$ densities can be computed beforehand for each
$i$ and $k$, and the conditional posterior distribution of $\gamma_i$
may be calculated directly over the simplex $\{0, \ldots, n\}$. 

When implementing this model, we restrict our data set to those firms
which have at least 20 observations, leaving us with 6067 data points.  Including firms with fewer data points would eliminate the possibility of survivorship bias, but would likely result in a significant decrease in power for the firms with longer histories, because of shared dependence of all firms on $p(\omega \mid \mathbf{Y})$.

Following the discussion above, the posterior distribution was
simulated using a Gibbs sampler with 3000 steps.  We set $\tau^2 =
10.0$ and chose the prior distributions as follows: $\dist{IG}(2.0 / 2, 2.0 / 2)$ for all noise variance parameters; and
$\bs{\omega} \sim \dist{Dirichlet}(\alpha)$ with $\alpha_0 = 0.8$, $\alpha_n = 0.1$,
and $\alpha_i = 0.1 / (n-1)$ for $i = 1, \ldots, n-1$.  This reflects the belief, borne out by prior studies, that most firms do not systematically over- or under-perform their peer groups over time.  The choice of $\tau^2 = 10$ will result in increased power for detecting major shifts in performance, but also a significant Occam's-razor penalty against small shifts.

One must be careful in interpreting the posterior distribution
$p(\bs{\gamma} | \bd{Y})$ that arises from this model.  For instance, it may not be meaningful to
categorize time series $i$ in terms of the single time point $j$ that maximizes $p(\gamma_i = j
| \bd{Y})$, since it is entirely possible that $p(\gamma_i = j | \bd{Y})$ is of comparable magnitude for a range of different values of $j$.  This would suggest either no split, or a split that cannot be localized very precisely.  In such cases, it may be more appropriate to look
at firms where the largest entry $p(\gamma_i \mid \mathbf{Y})$ is sufficiently large.  This would be strong evidence that there is a split at time $j$ for time series $i$.

\begin{figure}
\begin{center}
\includegraphics[width=2.7in]{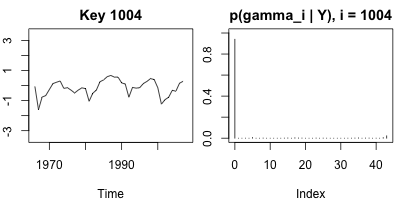} \\
\includegraphics[width=2.7in]{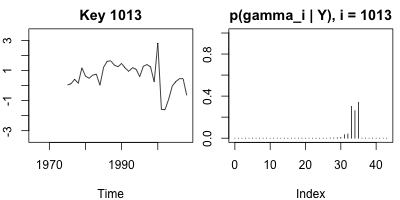} \\
\includegraphics[width=2.7in]{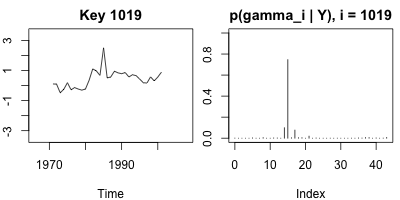} \\
\includegraphics[width=2.7in]{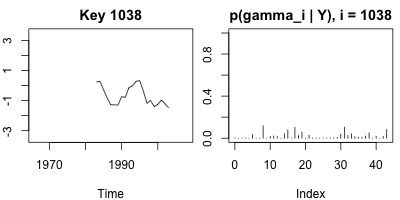} \\
\includegraphics[width=2.7in]{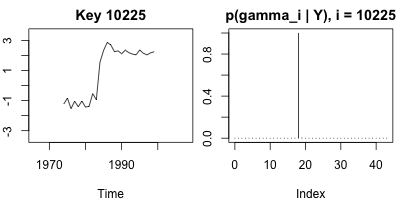}
\end{center}
\caption{\label{fig:five-examples}Five examples of a firm-level posterior distribution over possible changepoints, labeled by Compustat keys.  The five panes embody a wide range of possible conclusions, with the data on the left and $p(\gamma_i \mid \mathbf{Y})$ on the right.  From top to bottom, we see: strong evidence of a null case ($\gamma=0$); evidence of a hard-to-localize changepoint; moderate evidence of a specific changepoint; near-total uncertainty over changepoints; and very strong evidence of a specific changepoint.}
\end{figure}

Figure \ref{fig:five-examples}  provides intuition for the various
scenarios we might encounter: the posterior of $\gamma_i$ may strongly
favor $\gamma_i = 0$ or $\gamma_i = n$, in which case no split occurs;
it may show evidence of a split, but at an ambiguous time; it may show
strong, but not decisive evidence of a split; or it may be flat,
which does not tell us anything.  The ideal case for interpretation, of course, would involve strong
evidence of a split at a particular time, as seen in the last pane of
Figure \ref{fig:five-examples}.

\begin{figure}
\begin{center}
\includegraphics[scale=0.5]{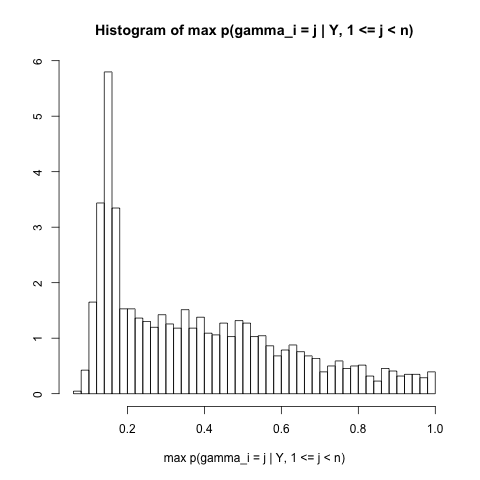}
\includegraphics[scale=0.5]{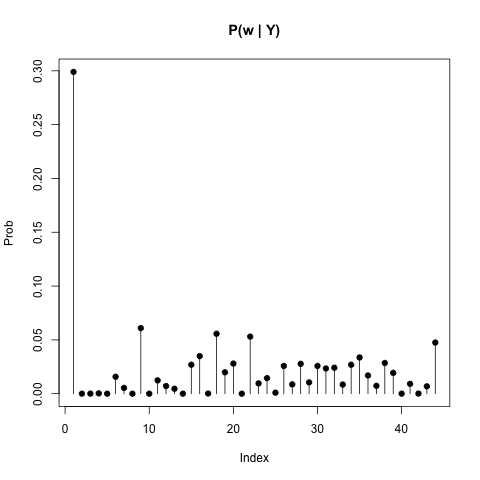}
\end{center}
\caption{\label{fig:posteriors}Top: the histogram of the largest entry in the posterior mode $PM(\gamma_i)$ across all firms.  Bottom: the posterior mean of the multinomial probability vector $\omega$.  Notice that null cases ($\gamma_i = 0$) dominate the sample, but that some years clearly have more changepoints than others.}
\end{figure}

In examining these plots, it appears that the posterior mode
\[
PM(\gamma_i) = \max_{0 < j < n} p(\gamma_i = j | \bd{Y})
\]
provides a good measure of a change in epochs as long as we chose a
sufficiently high cutoff, such as requiring that the maximum satisfy $\max_{0 < j < n} \{ \gamma_i = j |
\bd{Y} \} > 0.95$.  To check that this is reasonable, we plot the
histogram of $PM(\gamma_i)$ in Figure \ref{fig:posteriors}. 
Of the 3033 firms, only 58 have a value greater than or equal to
$0.95$.  The largest twenty posterior modes were used to select the time series in Figure \ref{fig:top20}.

\begin{figure}
\begin{center}
\label{fig:top20}
\includegraphics[angle=90,width=4.5in]{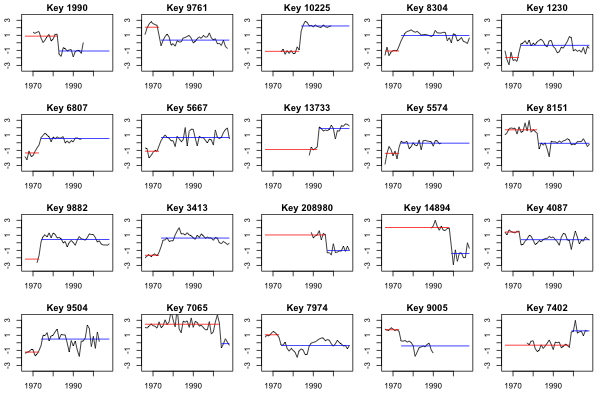}
\end{center}
\caption{\label{fig:top20}The firms with the 20 largest values for $PM(\gamma_i \mid \mathbf{Y})$.  These are the 20 firms in the cohort where the evidence for a specific changepoint is the strongest.}
\end{figure}

\section{Discussion}

In each of these two case studies, we have confronted a similar problem: time-varying model uncertainty for each of many different time series observed in parallel.  In the first case, the model was a graph, encoding conditional independence relationships about residuals from country-level returns during the European sovereign debt crisis.  In the second case, the model was an indicator of whether a firm's historical ROA trajectory was significantly different from its peer-group average.  These were seen to be examples of a general class of problems wherein the model changes as a function of auxiliary information.

The models we have entertained are based upon fairly standard tools, and were chosen specifically to avoid the difficulties associated with the most general form of partitioning that were mentioned in the introduction.  They are thus more appropriate for first-pass screening than for detailed analysis.  Nonetheless, even these simple models were sufficient to support the general thrust of our argument: that each data set exhibited non-trivial dynamic model structure.  In each case, further work is clearly needed to build upon the limited, broad-brush conclusions that can be reached within the context of these simple models.

\end{spacing}

\newpage

\begin{spacing}{1.0}
\begin{small}
\bibliographystyle{abbrvnat}
\bibliography{masterbib,business_articles}
\end{small}
\end{spacing}

\end{document}